\documentclass[twocolumn,letterpaper]{IEEEAerospaceCLS}  

\usepackage[]{graphicx}    
\newcommand{\ignore}[1]{}  

\usepackage{changes}
\usepackage{float}
\usepackage[justification=centering]{caption}
\usepackage{subcaption}
\usepackage{enumitem}
\usepackage[percent]{overpic}
\usepackage[ruled]{algorithm2e}
\usepackage[acronym, nonumberlist]{glossaries}
\usepackage{amsmath}
\usepackage{amssymb}
\usepackage{verbatim}
\usepackage[hidelinks]{hyperref}
\usepackage{url}

\DeclareMathOperator*{\argmin}{arg\,min}

\begin{document}
\title{Artificial Potential Field-Based Path Planning for Cluttered Environments}

\author{%
Mosab Diab\\
SPS, Faculty of EEMCS, Mekelweg 4 \\
Delft University of Technology (TUD)\\
2628CD Delft, The Netherlands
\and 
Mostafa Mohammadkarimi\\
SPS, Faculty of EEMCS, Mekelweg 4 \\
Delft University of Technology (TUD)\\
2628CD Delft, The Netherlands
\and
Raj Thilak Rajan\\
SPS, Faculty of EEMCS, Mekelweg 4 \\
Delft University of Technology (TUD)\\
2628CD Delft, The Netherlands
\thanks{\footnotesize This work was published at the 2023 IEEE Aerospace Conference under the DOI: 10.1109/AERO55745.2023.10115857}              
}

\maketitle

\thispagestyle{plain}
\pagestyle{plain}

\maketitle

\thispagestyle{plain}
\pagestyle{plain}

\begin{abstract}
In this paper, we study path planning algorithms of resource constrained mobile agents in unknown cluttered environments, which include but are not limited to various terrestrial missions e.g., search and rescue missions by drones in jungles, and space missions e.g., navigation of rovers on the Moon. In particular, we focus our attention on artificial potential field (APF) based methods, in which the target is attractive while the obstacles are repulsive to the mobile agent. In this paper, we propose two major updates to the classical APF algorithm which significantly improve the performance of path planning using APF. First, we propose to improve an existing classical method that replaces the gradient descent optimization of the potential field cost function on a continuous domain with a combinatorial optimization on a set of predefined points (called bacteria points) around the agent’s current location. Our proposition includes an adaptive hyperparameter that changes the value of the  potential function associated to each bacteria point  based on the current environmental measurements. Our proposed solution improves the navigation performance in terms of convergence to the target at the expense of minimal increase in computational complexity. Second, we propose an improved potential field cost function of the bacteria points by introducing a new branching cost function which further improves the navigation performance. The algorithms were tested on a set of Monte Carlo simulation trials where the environment changes for each trial. Our simulation results show 25\% lower navigation time and around 300\% higher success rate compared to the conventional potential field method, and we present future directions for research.
\end{abstract}

\tableofcontents

\section{Introduction}

Autonomous agents such as drones and rovers are the future of delivery systems, search and rescue missions, and surveillance systems. Autonomous agents can be used in manufacturing plants, such as pick and place robots. They can also be used for disinfection in medical facilities. Moreover, they can be used in transport of both people and goods and many more different applications.

The navigation systems for drones in unknown and cluttered environments like jungles, oceans, and sea floors are in their early development stages and still do not have the sufficient performance levels to be deployed into the field. Specifically, low complexity 
path planning and collision avoidance algorithms are needed to be developed for autonomous agents in cluttered environments. Path planning is defined as finding a geometrical path from the current location of the agent to the target location such that it avoids the obstacles in the environment \cite{Azadi2021}. The goal is to reach the target location through a safe path in the shortest time possible.


In this paper, we propose a path planning algorithm optimized for cluttered environments that outperforms existing classical path planning algorithms in terms of convergence and time it takes to navigate. We first start by analyzing the existing methods and deciding which one is the most suitable for use in cluttered environments. We then define our system model and describe the artificial potential field (APF) method. After that, the path planning system developed for cluttered environments is described along with the contributions we made. We then discuss the local minima trap which is the biggest drawback of APF methods. We conclude the paper with the simulation setup and results.

\subsection{Related Research}

The working principle of existing collision avoidance systems employed for path planning can be explained by either reactive control or deliberative planning. Reactive control is when the agent gathers information using on-board sensors and reacts based on the gathered data. Reactive control allows for rapid response but can lead to a corner case in the path and might get the agent stuck so it may need another technique combined with it in order to avoid that. Deliberative planning on the other hand is when there is an environmental map that is constantly updated by the agent. Then, the optimal collision free navigation path is calculated. The latter method needs an accurate map of the environment and that is computationally extensive specially for a dynamic environment. A hybrid approach between the two is considered more suitable for dynamic environments \cite{Yasin2020}. 

Existing collision avoidance methods can be divided into four groups:
\begin{enumerate}[leftmargin=0.7cm,noitemsep,partopsep=0pt,topsep=0pt,parsep=0pt]
    \item Geometric Methods
    \item Force-Field Methods
    \item Optimisation-Based Methods
    \item Sense-and-Avoid Methods
\end{enumerate}
\subsubsection{Comparison Between Methods}

Multiple parameters can be used for performance comparison of collision avoidance methods employed for path planning. In this section, we will review some of these methods and try to indicate the discrepancy.

The real time performance of the sense and avoid and geometric methods is better than the force-field and optimisation methods \cite[p. 11]{Yasin2020}. Sense and avoid does not increase computational complexity if a change happens in the environment such as an obstacle moving \cite{Gageik2015}. Geometric methods are also computationally light but are highly dependent on the algorithm implementation in terms of computation time \cite{Seo2017}. Optimisation methods are of medium complexity while the force-field methods are considered low-complex in implementation \cite[p. 11]{Yasin2020}.  The velocity constraint is a metric that takes the velocity of the obstacles in consideration. Sense and avoid and geometric approaches are capable to handle this constraint well \cite{Wang2018},\cite{Bareiss2013}. However, force-field and optimisation methods are better for predefined planning which does not take changing dynamics into consideration \cite[p. 11]{Yasin2020}. The third metric is static and dynamic environment suitability. For dynamic environments, sense and avoid is the easiest and lightest due to the local computations being performed on board the agent through the changes observed by the sensory system on the agent itself \cite{Hrabar2011}. Geometric methods give an acceptable performance but less optimal than sense and avoid \cite{Conroy2014}. Force-field methods do not perform well in narrow passages and can lead to a local minima trap in dynamic environments \cite{Roelofsen2015}. Optimisation methods are more suitable for static environments as they require pre-planning and have to optimise the whole route again in case a change is detected \cite[pp. 11-12]{Yasin2020}.

When it comes to deadlock (local minima trap), the geometrical and optimisation methods do not have this issue due to the fact that they are designed for known structured environments. Force-field methods can get stuck in a local minima. Sense and avoid methods can reach a deadlock and require another methodology for handling this issue as it cannot be solved locally \cite[p. 12]{Yasin2020}. When it comes to swarm compatibility all the approaches can be applied to a swarm of agents but might need an additional algorithm for handling communication between the agents. When it comes to dimensionality, all the mentioned methods need some minor modifications when it comes to scaling the algorithms developed for 2D to handle 3D environments. However, there is current research on the feasibility of force-field methods for 3D dynamic environments \cite{Yasin2020},\cite{Sun2017}. For pre-mission path planning, sense and avoid and force field methods do not require that because the plan is made when the obstacle is detected in-flight. Optimisation and geometric methods on the other hand, require path planning beforehand as the obstacle locations need to be known to the UAV before encountering the obstacles.

The takeaway point here is that for unknown cluttered environments, the force-field method is the best one to use. The reason for that is that the geometric and optimisation methods are designed for known structured environments and do not perform well in unknown cluttered environments while the sense and avoid methods do not take multiple obstacles into account which is necessary for cluttered environments.


\section{System Model}


We consider an agent with a known position ${\bf r}=[{x},{y}]^T$ which aims to navigate through a set of ${N}$ cluttered static obstacles in a 2D environment\footnote{Extension to 3D case is straightforward.} to a target of a known location ${\bf r}_{\rm t}=[{x}_{\rm t},{y}_{\rm t}]^T$.
The agent uses dead reckoning technique to obtain its location during navigation \cite{Randell2005}. It is assumed that the number and location of the obstacles are unknown to the agent.
The location of the obstacles are considered to be a random vector uniformly distributed in a rectangular environment with length $L_{\rm x}$ and $L_{\rm y}$. 
The agent starts the navigation from a known location ${\bf r}_{\rm s}=[{x}_{\rm s},{y}_{\rm s}]^T$ and it is equipped with a range measurement sensor that measures the distances to the obstacles.


The detection range of the agent's sensor is denoted by $\rho_{\rm rn}$. The agent can detect only the obstacles within its detection range in each navigation step. Thus, the number of detected obstacles varies during navigation. 
The navigation is composed of a measurement step followed by path planning, where it is finding a geometrical path from the current location of the agent to its next location while avoiding obstacles in the environment.

\begin{figure*}

\centering
     \begin{subfigure}[c]{0.45\textwidth}
         \centering
         \includegraphics[width=\textwidth]{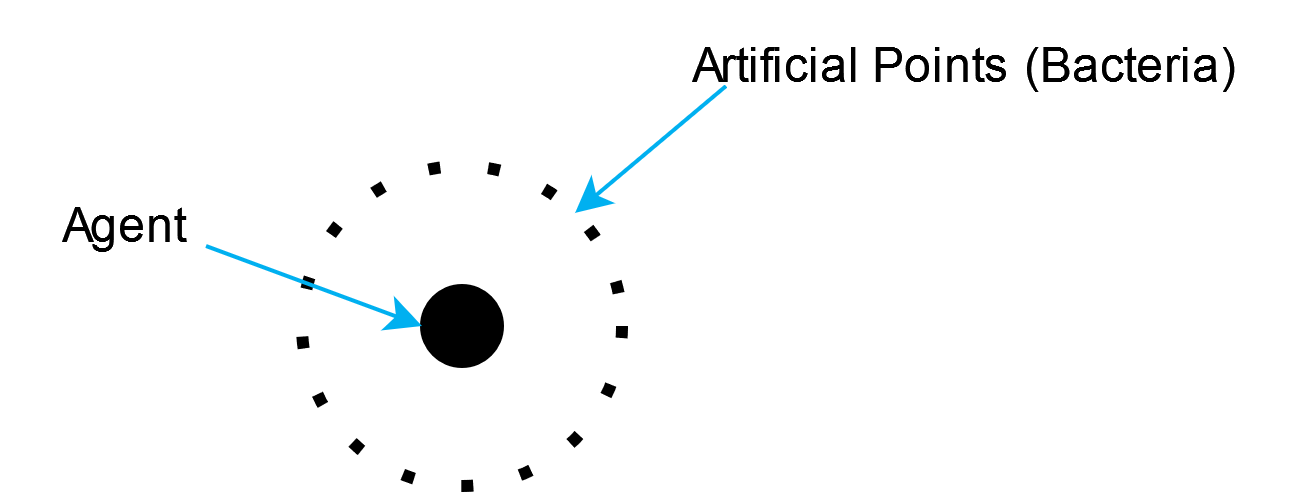}
         \label{fig:bacteria_points}
     \end{subfigure}
     \begin{subfigure}[c]{0.52\textwidth}
         \centering
         \includegraphics[width=\textwidth]{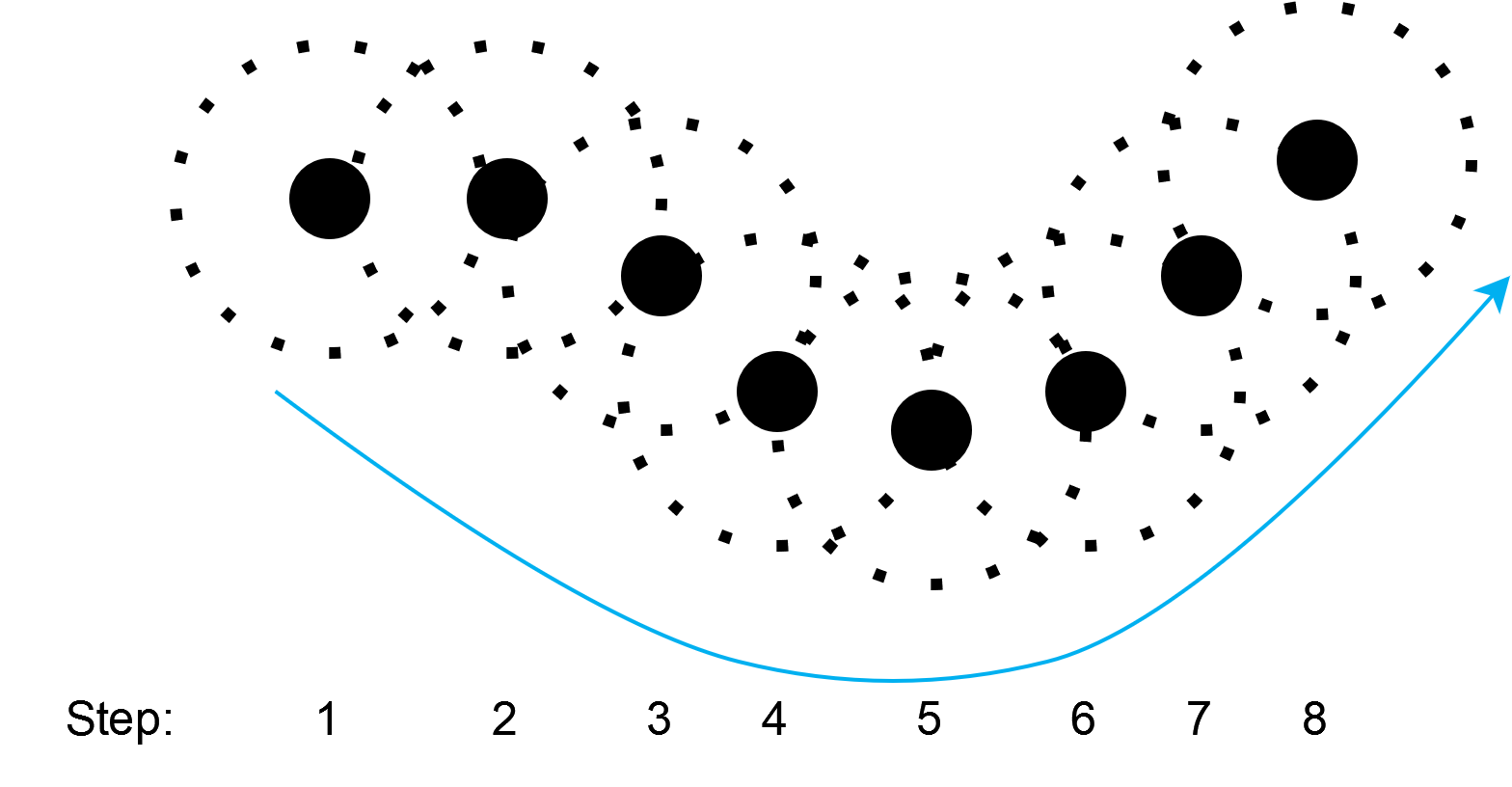}
         \label{fig:agent_movement}
     \end{subfigure}
        \caption{Agent and bacteria points around it (left) Agent movement by selecting different bacteria points (right).}
        \label{fig:bacteria_illustration}

\end{figure*}

\section{Artificial Potential Fields (APF)}  \label{APF}
In this section, we briefly overview classical APF methods for unknown cluttered environments. 
In general, the APF methods works on the basis that a target is attractive while an obstacle is repulsive. The agent is to be attracted by the goal point and hence moves towards it while changing direction whenever an obstacle is encountered. An example for this is a mobile electrical charge moving in a field of static similar charges while being attracted by a static opposite charge. The advantage of this method is that it takes into account the effect of multiple obstacles at once which makes it perfect for cluttered environments which are the scope of this research.

\subsection{Classical APF (CAPF) Algorithm}

The original APF method proposed in 1985 \cite{Khatib1985} uses geometric calculations at each iteration to determine the direction and speed the agent should move on in order to avoid the obstacles and reach the target safely. It was originally proposed for robotic arm manipulators but has been extended to mobile agents afterwards \cite{Mohanta2016}.

The CAPF method was originally designed for structured environments. It can still work in cluttered environments but is far from optimal and often gets stuck in local minimas and collides with obstacles. Hence, the need arises for a new method that is dedicated for cluttered environments and that is the purpose of this paper.

\subsection{Bacteria APF (BAPF) Algorithm}

The principle idea of the BAPF method for autonomous navigation is illustrated in Figure \ref{fig:bacteria_illustration} and is adopted from \cite{Ahmad2020}. As seen, it is based on the idea of having a point agent surrounded by a circle of possible future position points called bacteria points. There are potential functions that are calculated for those points based on the distance from the target and from the detected obstacles.  Based on those potential functions a certain bacteria point is selected to be the next position of the agent, and the agent moves to it. This process is repeated during navigation \cite{Ahmad2020}. This method shall overcome the limitations that the CAPF method faces when it comes to unknown cluttered environments in terms of the agent getting stuck between the scattered obstacles \cite{Mohanta2016}. 



\begin{figure*}[!htbp]
    \centering
    \begin{overpic}[scale=0.75]{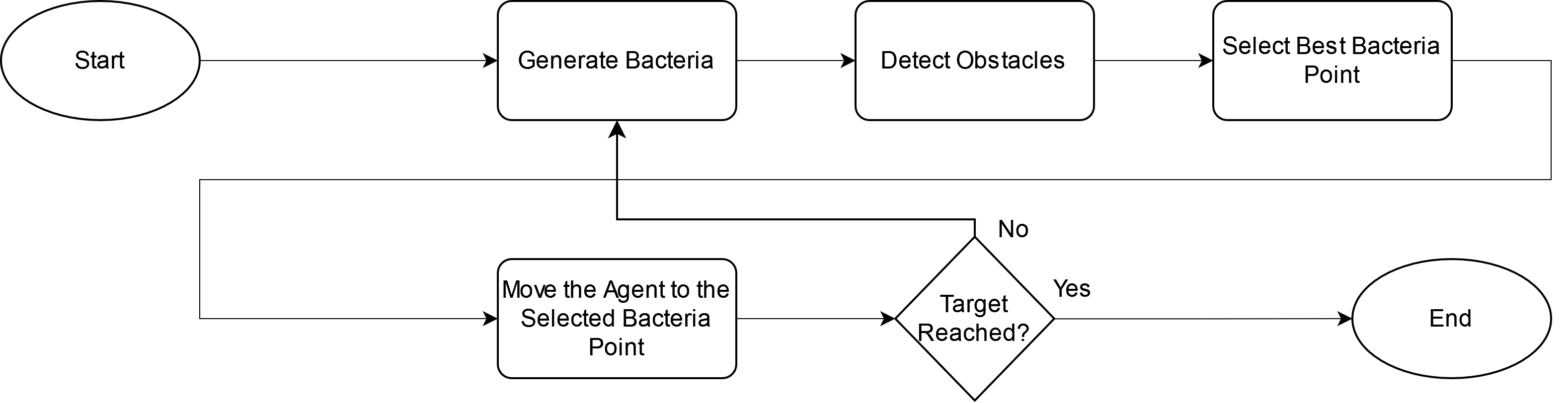}
    \put (81.2,18.7) {\footnotesize\eqref{eq:total_agent_potential},\eqref{eq:total_bacteria_potential},\eqref{eq:condition}}
    \end{overpic}
    \caption{BAPF algorithm flow graph for agent navigation.}
    \label{fig:flow_graph}
\end{figure*}

\begin{figure*}[!htbp]
    \centering
    \begin{overpic}[scale=0.7]{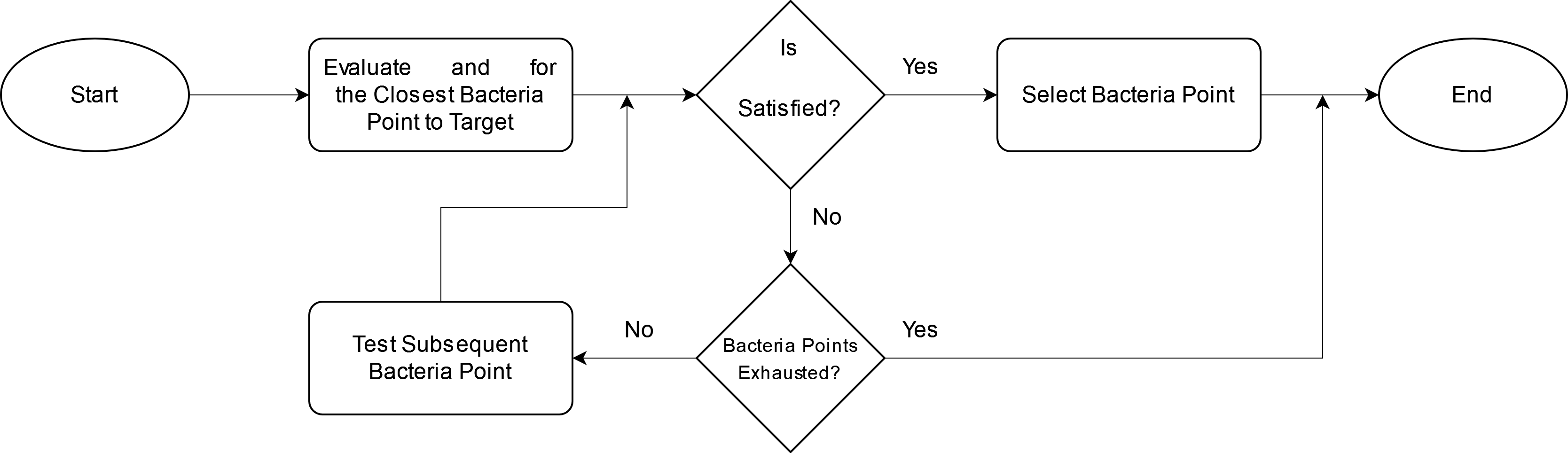}
    \put (26.4,24) {\footnotesize\eqref{eq:total_agent_potential}}
    \put (31.4,24) {\footnotesize\eqref{eq:total_bacteria_potential}}
    \put (49.3,23.5) {\footnotesize(\ref{eq:condition})}
    \end{overpic}
    \caption{BAPF algorithm flow graph for best bacteria point selection.}
    \label{fig:flow_graph2}
\end{figure*}

In the BAPF algorithm,  the potential function from the target located at ${\bf r}_{\rm t}=[{x}_{\rm t},{y}_{\rm t}]^T$ to the agent located at ${\bf r}=[{x},{y}]^T$ is given by 
\begin{equation}
    J_{\rm t}(\textbf{r})\triangleq-\alpha_{\rm t}\exp(-\mu_{\rm t}\rho_{\rm t}(\textbf{r})),
    \label{eq:target_potential}
\end{equation}
where $\rho_{\rm t}({\bf r}) \triangleq \lVert\textbf{r}-\textbf{r}_{\rm t}\rVert_{2}^2$ is the square distance between the agent and target, and $\alpha_{\rm t}$ and $\mu_{\rm t}$ are constant parameters empirically determined based on the size of the environment and the initial distance between the agent and the target. 

Similarly, the repulsive potential function from the $n$th detected obstacle located at ${\bf r}_{\rm o}^n=[{x}_{\rm o}^n,{y}_{\rm o}^n]^T$ to the agent located at ${\bf r}=[{x},{y}]^T$ is defined as
\begin{equation}
    J^{n}_{\rm o}(\textbf{r})\triangleq\alpha_{\rm o}\exp(-\mu_{\rm o} \rho_{\rm o}^n({\bf r})),
    \label{eq:repulsive_potential}
\end{equation}
where $\rho_{{\rm o}}^n(\textbf{r}) \triangleq \lVert\textbf{r}-\textbf{r}_{\rm o}^n\rVert_{2}^2$ is the square distance between the agent and the $n$th obstacle, and $\alpha_{\rm o}$ and $\mu_{\rm o}$ are constant parameters determined through experimentation based on the estimated density of obstacles in the navigation region.

Let $N_{\rm d}$ denote the total number of detected obstacles by the agent at the location of  ${\bf r}=[{x},{y}]^T$. The total repulsive potential function from the $N_{\rm d}$ detected obstacles at the location of the agent  ${\bf r}=[{x},{y}]^T$ is given by
\begin{equation}\label{total}
    J_{\rm o}(\textbf{r})=\sum_{n=1}^{N_{\rm d}}J^{n}_{\rm o}(\textbf{r}), 
\end{equation}
where $J^{n}_{\rm o}(\textbf{r})$ is given in \eqref{eq:repulsive_potential}. 
By using \eqref{eq:target_potential} and \eqref{total}, the total potential function at the agent's location
${\bf r}=[{x},{y}]^T$ can be written as
\begin{align}
J_{\rm a}(\textbf{r})=J_{\rm t}(\textbf{r})+J_{\rm o}(\textbf{r}).
\label{eq:total_agent_potential}
\end{align}

Let $N_{\rm b}$ and ${\bf r}_{{\rm b},k}=[{x}_{{\rm b},k},{y}_{{\rm b},k}]^T$, $k=1,2,\ldots,{N_{\rm b}}$, denote the total number of bacteria points and the location of the $k$th bacteria point around ${\bf r}=[{x},{y}]^T$ respectively.  
The total potential function at the location of the $k$th bacteria point around ${\bf r}=[{x},{y}]^T$ is expressed as 
\begin{align}\label{eq:total_bacteria_potential}
J_{k}(\textbf{r}_{{\rm b},k})=J_{\rm t}(\textbf{r}_{{\rm b},k})+J_{\rm o}(\textbf{r}_{{\rm b},k}).
\end{align}

The criterion for the movement of the agent located at ${\bf r}=[{x},{y}]^T$
to the $k$th bacteria point at ${\bf r}_{{\rm b},k}=[{x}_{{\rm b},k},{y}_{{\rm b},k}]^T$ is given by

\begin{equation}
    J_{k}({\bf r}_{{\rm b},k})-J_{\rm a}(\textbf{r})<0, 
    \label{eq:condition}
\end{equation}
which  means that the $k$th bacteria point is at a lower total potential towards the target than the agent. Recall from \eqref{eq:target_potential} and \eqref{eq:repulsive_potential} that the target potential is negative while the repulsive potential from the obstacles is positive.

Figure \ref{fig:flow_graph} illustrates the flow graph of the BAPF algorithm. As seen,  a linear search algorithm on the bacteria points is performed to select the best bacteria point to move to. The flow graph of the search algorithm is shown in Figure \ref{fig:flow_graph2}. 
As seen, the search starts with testing the bacteria point closest to the target. If that point meets the potential criteria in \eqref{eq:condition}, then the agent moves to it without testing other bacteria points. If the bacteria point does not meet the potential criteria then the second closest bacteria point to the target is tested. This procedure is repeated until a bacteria point meets the potential criteria in \eqref{eq:condition}. When a bacteria point is selected it becomes the next point that the agent should move to. The distance between the agent and the bacteria points is constant and is defined as the movement step size of the agent. It is determined empirically given the type of environment and the detection range of the agent's sensor.


\subsubsection{Local Minima Trap}

The major drawback of the APF methods is that they can get stuck in a local minima where the agent cannot find a solution to the problem on a certain area on the navigation region. The local minima trap is sometimes inevitable in the case of unknown environments and especially unknown cluttered environments.
For the BAPF algorithm, the local minima trap occurs if none of the $N_{\rm b}$ bacteria points meets the criteria in \eqref{eq:condition}. In this case, the agent cannot move.

\begin{figure}
         \centering
         \includegraphics[width=0.5\textwidth]{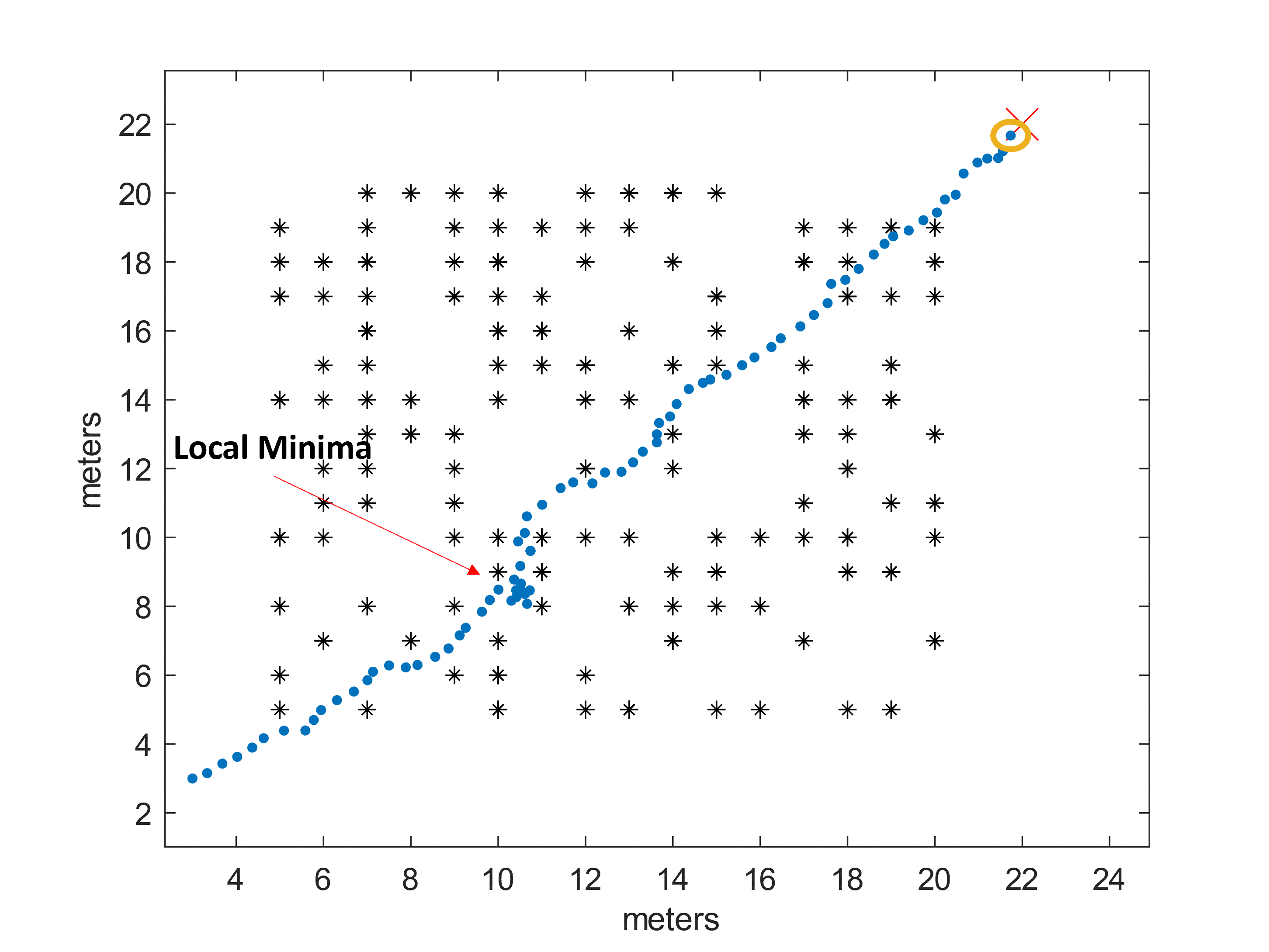}
        \caption{Local minima escape through random walk}
        \label{fig:local minima escape}
\end{figure}

To escape the local minima trap, a random walk can be employed where a bacteria point is selected randomly from the $N_{\rm b}$ bacteria points, and the agent to move to it \cite{Xia2019}. A strict condition for this randomly selected bacteria point is that it should not violate a minimum safety distance to a detected obstacle which will be further discussed later on. Let $q \in \{1,2,\ldots,N_{\rm b}\}$, denote the index of the selected bacteria point for the next movement through the random walk. The mathematical formulation of the random walk movement can be expressed as
\begin{align}
 q \sim {\mathcal U}\{\mathcal Q\},
\end{align}
where 
\begin{align}
 {\mathcal Q} = \Big{\{}k \in \{1,2,\ldots, N_{\rm b}\}\,|\,\sqrt{\rho_{{\rm o}}^{n}(\textbf{r}_{{\rm b},k})}\geqslant \rho_{\rm l}\Big{\}}
\end{align}
and 
${\mathcal U}\{\mathcal Q\}$ returns an element of set ${\mathcal Q}$ randomly, $\sqrt{\rho_{{\rm o}}^{n}(\textbf{r}_{{\rm b},k})} = \lVert\textbf{r}_{{\rm b},k}-\textbf{r}_{\rm o}^n\rVert_{2}$, $N_{\rm b}$ is the number of bacteria points, and $\rho_{\rm l}$ denotes the safety distance as it is explained in the next section. Figure \ref{fig:local minima escape} shows the agent escaping a local minima trap via random walk.

\section{Improved APF Algorithms}

In this section, we propose the Adaptive BAPF (A-BAPF) and the Changing Radii BAPF (CR-BAPF) algorithms to improve the performance of the BAPF algorithm for unknown cluttered environments.

\subsection{A-BAPF Algorithm}

The BAPF algorithm cannot achieve the best performance in unknown cluttered environments when it comes to the convergence to the target (successful navigation). Performance improvement in terms of convergence to the target  can be achieved by employing our proposed A-BAPF method that is explained in this section.

The idea of A-BAPF is to adaptively  choose the value of 
 $\mu_{\rm o}$ for each bacteria point given a predetermined value of $\alpha_{\rm o}$. We define the modified total bacteria function as
 \begin{align}
 \bar{J}_{k}(\textbf{r}_{{\rm b},k},\mu_{{\rm o},k})
 \triangleq
 J_{\rm t}(\textbf{r}_{{\rm b},k})+\bar{J}_{\rm o}(\textbf{r}_{{\rm b},k},\mu_{{\rm o},k}), 
 \end{align}
 where 
 \begin{align} \label{eq:repulsive_A-BAPF}
 \bar{J}_{\rm o}(\textbf{r}_{{\rm b},k},\mu_{{\rm o},k})
 \triangleq\sum_{n=1}^{N_{\rm d}}\alpha_{{\rm o}}\exp(-\mu_{{\rm o},k} \rho_{\rm o}^n({\bf r_{\rm b,k}})), 
 \end{align}
 and 
\begin{equation}\label{eq:potential_optimization}
\begin{array}{r c c c}
 \hat{\mu}_{{\rm o},k}= \displaystyle   \argmin_{\mu_{{\rm o},k}} & {\bar J}_{k}({\bf r}_{{\rm b},k},\mu_{{\rm o},k}) \\
    \textrm{subject to} & {\mu_{{\rm o},k}} \in [\mu_{\rm min}:\mu_{\rm max}]
\end{array}
\end{equation}
 for $k=1,2,\ldots,N_{\rm b }$, 
and $\mu_{\rm min}$ and $\mu_{\rm max}$ are the lower and upper values of $\mu_{{\rm o},k}$.
Now, the criterion for the movement in \eqref{eq:condition} is replaced by 
\begin{equation}
    \bar{J}_{k}(\textbf{r}_{{\rm b},k},\hat{\mu}_{{\rm o},k})-J_{\rm a}(\textbf{r})<0. 
    \label{eq:A-BAPF_condition}
\end{equation}

One can easily show that ${\bar J}_{k}({\bf r}_{{\rm b},k},\mu_{{\rm o},k})$ in \eqref{eq:potential_optimization} is a convex function of  $\mu_{{\rm o},k}$. The solution of the minimization in \eqref{eq:potential_optimization} can be obtained by using grid search or golden-section search methods.
Intuitively,  when the value of $\mu_{{\rm o},k}$  approaches $\mu_{\rm max}$, the repulsive potential, $\bar{J}_{\rm o}(\textbf{r}_{{\rm b},k},\mu_{{\rm o},k})$ in \eqref{eq:repulsive_A-BAPF}, approaches zero and the obstacles' influence on the navigation decreases. The minimization of the potential cost function for the bacteria points can almost guarantee a solution for \eqref{eq:A-BAPF_condition} at each navigation step and prevent the agent from ending up in a local minima. This is due to the fact that the optimization in \eqref{eq:potential_optimization}  increases the chance of that at least one bacteria point 
 meets the condition in \eqref{eq:A-BAPF_condition}. 
Our proposed A-BAPF algorithm can be considered as an upper bound for the BAPF algorithm in terms of successful navigation performance. In the A-BAPF algorithm, the bacteria point closest to the target can be selected even if it does not initially meet the condition in \eqref{eq:condition}. This is done by changing the value of $\mu_{{\rm o},k}$ through \eqref{eq:potential_optimization}.



\subsection{CR-BAPF Algorithm} \label{sec:Potential Changing Radii}
In the previously discussed APF methods, the potential from a detected obstacle is always affecting the agent during navigation. In this section, we propose the CR-BAPF algorithm where the repulsive potentials from obstacles affecting the agent change based on regions around each obstacle.

\begin{figure}[!t]
    \centering
    \includegraphics[width=0.5\textwidth]{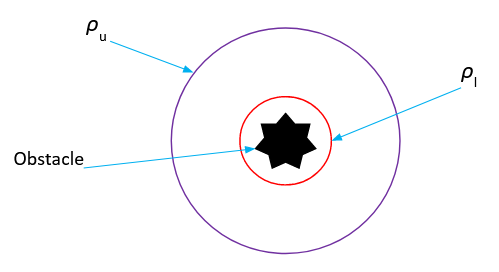}
    \caption{The upper and lower radii around a point obstacle.}
    \label{fig:Radii_Graph}
\end{figure}

The CR-BAPF algorithm introduces two radii around each obstacle as shown in Figure \ref{fig:Radii_Graph}. For the CR-BAPF algorithm, we define the repulsive potential function from the $n$th detected obstacle located at ${\bf r}_{\rm o}^n=[{x}_{\rm o}^n,{y}_{\rm o}^n]^T$, $n=1,2,\ldots,N_{\rm d}$, to the agent located at ${\bf r}=[{x},{y}]^T$ as
\begin{equation}
\scriptstyle
  J^{n}_{\rm o}(\textbf{r})=\begin{cases}
    0 & \text{ $\sqrt{\rho^{n}_{\rm o}(\textbf{r})} \textgreater \rho_{\rm u}$}\\
    \alpha_{\rm o}\exp(-\mu_{\rm o}\rho_{{\rm o}}^n(\textbf{r})) & \text{ $\rho_{\rm l} \leqslant \sqrt{\rho^{n}_{\rm o}(\textbf{r})} \leqslant \rho_{\rm u}$}\\
    \infty & \text{  $\sqrt{\rho^{n}_{\rm o}(\textbf{r})} \textless \rho_{\rm l}$}
  \end{cases}, 
  \label{eq:branching_obstacle_potential}
\end{equation}
where $\rho_{{\rm o}}^n(\textbf{r}) \triangleq \lVert\textbf{r}-\textbf{r}_{\rm o}^n\rVert_{2}^2$ is the square distance between the agent and the $n$th obstacle, and
$\rho_{\rm l}$ and $\rho_{\rm u}$ denote the lower and upper radii, respectively.
As seen in \eqref{eq:branching_obstacle_potential},  if the agent's distance to the $n$th detected obstacle is greater than $\rho_{\rm u}$, then the repulsive potential is zero. On the other hand, if the agent's distance to the $n$th obstacle is lower than $\rho_{\rm l}
$, the repulsive potential from the obstacle to the agent becomes infinity. This means that the agent does not select a bacteria point that its distance to the $n$th obstacle is lower than $\rho_{\rm l}$ because it does not satisfy the criteria in (\ref{eq:condition}). Moreover, this also implies that the agent is not affected by obstacles that are at a greater distance than $\rho_{\rm u}$. This way, the agent has more motion freedom in the map without risking its safety which improves the overall performance. The lower radius is meant as a safety perimeter around the obstacles to prevent collisions.
The value of  $\rho_{\rm l}$ is determined based on the error margin in the motion of the agent in order to prevent the agent from colliding with the obstacle. Moreover, the value of $\rho_{\rm u}$ is determined based on the estimated density of the obstacles in the navigation region. It should be mentioned that the attractive potential function to the target for the CR-BAPF is given in \eqref{eq:target_potential}.

\subsection{CR-BAPF* Algorithm}
To further improve the performance of the proposed CR-BAPF algorithm in terms of avoiding the local minima trap discussed in \ref{APF}, we propose to employ the CR-BAPF algorithm in combination with the random walk technique. We call this new algorithm CR-BAPF\textsuperscript{*}.

\section{Simulations}

\subsection{Simulation Setup}

The simulation environment is setup in MATLAB with a fixed starting position for the agent at ${\bf r_{\rm s}}=[3,3]^T$ meters and the target location at ${\bf r}_{\rm t}=[22,22]^T$ meters. The location of the obstacles is considered to be a random
vector ${\bf r}_{\rm o}^n=[{x}_{\rm o}^n,{y}_{\rm o}^n]^T$, $n=1,2,\ldots, N$, uniformly distributed in a square region with length $L_{\rm x}=30$ m and width $L_{\rm y}=30$ m.  For each Monte Carlo trial, the number of obstacles in the navigation region $N$ is modeled by the discrete uniform random variable $N \in {\mathcal{U}}{[N_{\rm l},N_{\rm u}]}$, where $N_{\rm l}$ and $N_{\rm u}$ are the lower and upper limit for the number of obstacles, respectively. The considered simulation setup emulates the
navigation through a set of randomly uniformly distributed
obstacles such as trees or craters.
The detection range is set $\rho_{\rm rn} =8$ m, and the number of bacteria points is $N_{\rm b}=60$ 
to provide a rotation accuracy of $6^{\circ}$.

The movement step size of the agent is $\Delta r \triangleq \lVert\textbf{r}-\textbf{r}_{{\rm{b}},k} \rVert_{2}=0.4$ m, $k=1,2,\ldots,N_{\rm b}$, and 
the position errors of the agent $\Delta x$ and $\Delta y$ are modeled by a zero mean Gaussian distribution with variance $0.01$.
 The values of the hyperparameters for $N_{\rm m}=4000$ Monte Carlo trials are given  in Table \ref{table:Hyperparameters}. The upper  and lower radii are considered the same for all the obstacles in the navigation region.


\begin{table}[!t]
\renewcommand{\arraystretch}{1.4}
\caption{{\bf Simulation hyperparameter values.}}
\begin{center}
\begin{tabular}{ |p{0.285\linewidth}||p{0.14\linewidth}|} 
 \hline
 Hyperparameter & Value \\ 
 \hline\hline
\,\,\,\,\,\,\,\,\,\,\,\,\,\,\,\,\,\,\ $\rho_{\rm rn}$ & 8 m\\ 
 \hline
\,\,\,\,\,\,\,\,\,\,\,\,\,\,\,\,\,\,\ $\alpha_t$ & ${\rm 10^4}$\\
 \hline
\,\,\,\,\,\,\,\,\,\,\,\,\,\,\,\,\,\,\ $\mu_{\rm t}$& 1\\ 
 \hline
\,\,\,\,\,\,\,\,\,\,\,\,\,\,\,\,\,\,\ $\alpha_{\rm o}$& 1\\
 \hline
\,\,\,\,\,\,\,\,\,\,\,\,\,\,\,\,\,\,\ $\mu_{\rm o}$& 1000\\
 \hline
\,\,\,\,\,\,\,\,\,\,\,\,\,\,\,\,\,\,\  $\rho_{\rm l}$& 0.4 m\\
 \hline
\,\,\,\,\,\,\,\,\,\,\,\,\,\,\,\,\,\,\ $\rho_{\rm u}$& 4.5 m\\
 \hline
\end{tabular}
\end{center}

\label{table:Hyperparameters}
\end{table}

\begin{figure*}
     \centering
     \begin{subfigure}[c]{0.497\textwidth}
         \centering
         \includegraphics[width=\textwidth]{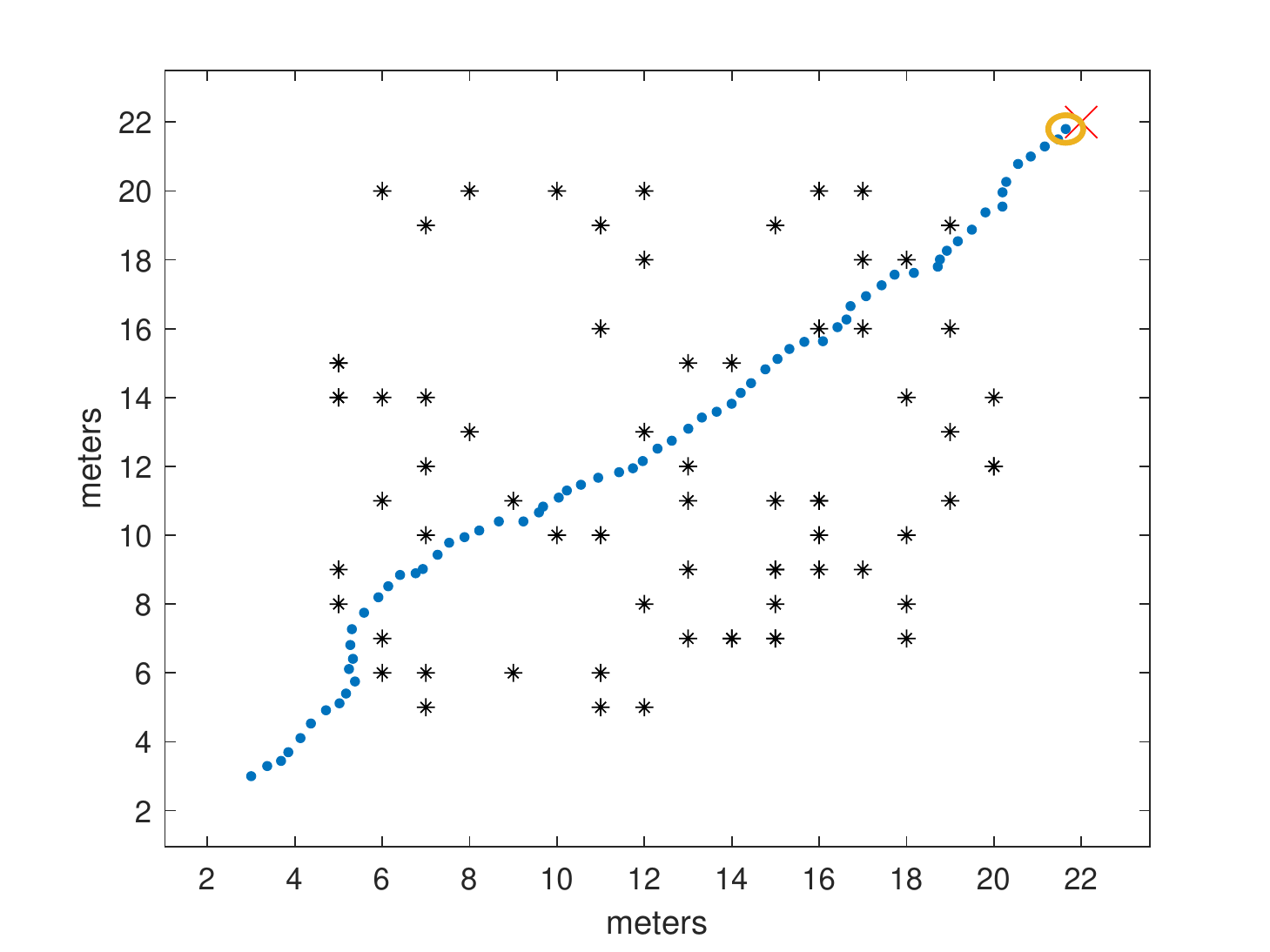}
         \caption{Successful navigation}
         \label{fig:Successful_Run}
     \end{subfigure}
     \begin{subfigure}[c]{0.497\textwidth}
         \centering
         \includegraphics[width=\textwidth]{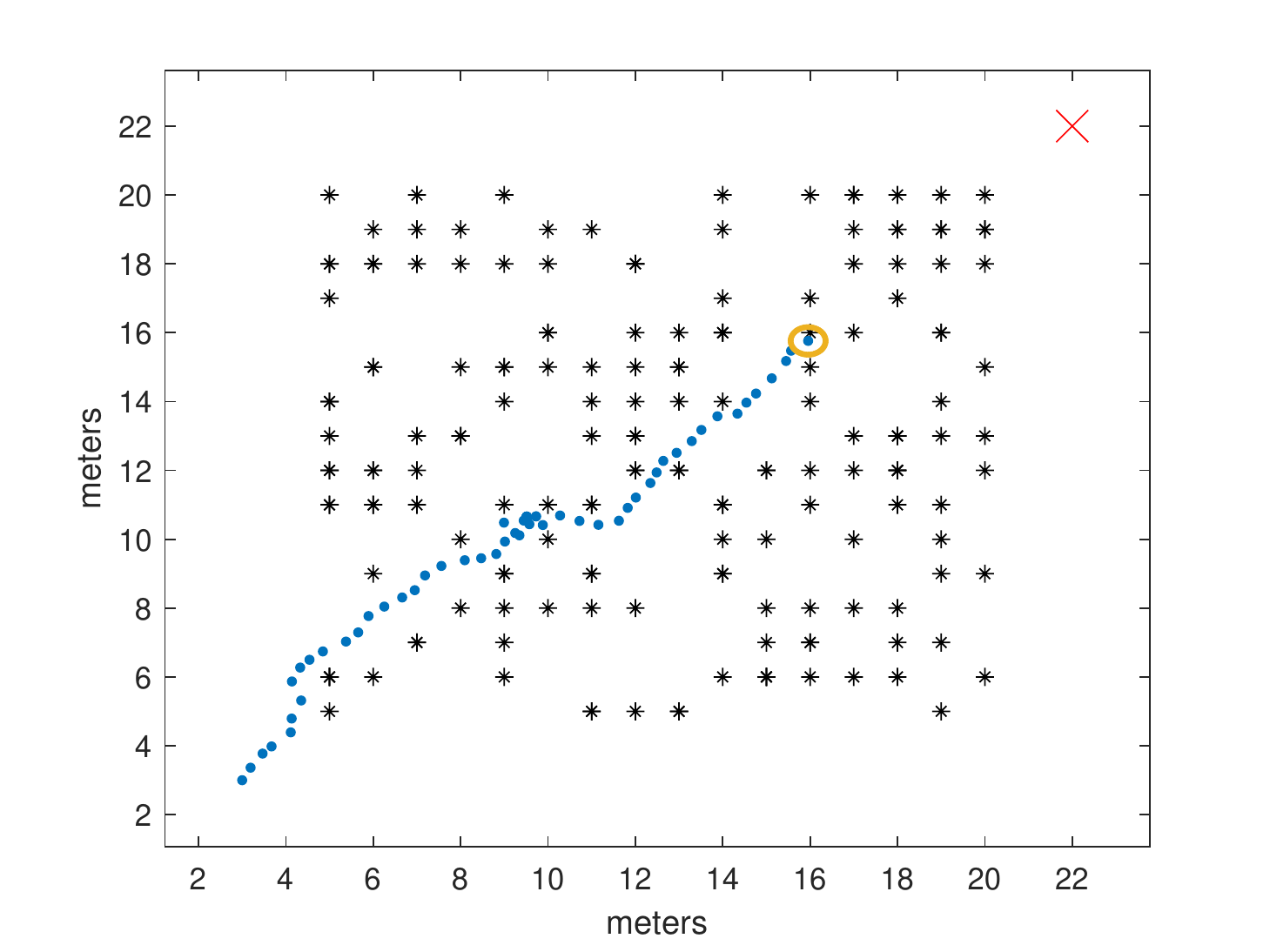}
         \caption{
Unsuccessful navigation}
         \label{fig:Getting_Stuck_in_Run}
     \end{subfigure}
        \caption{Successful and unsuccessful navigation of an agent
        employing CR-BAPF\textsuperscript{*} algorithm.}
        \label{fig:navigation_results}
\end{figure*}

\subsection{Performance Metrics} \label{Comparison_Criteria}

The following terms and definitions will be used in order to compare different algorithms.

\subsubsection{\it{Rate of Success} ($R_{\rm s}$)} For ${N_{\rm m}}$ Monte Carlo trials, the rate of successful navigation is defined as 
\begin{equation}
    R_{\rm s} \triangleq \frac{N_{\rm s}}{N_{\rm m}}
    \label{eq:Success_Rate}, 
\end{equation}
where 
${N_{\rm s}}$ denotes the number of successful navigation without getting stuck in a local minima or colliding with an obstacle. 


\subsubsection{Average Navigation Steps ($\bar{M}_{\rm s}$)} For $N_{\rm s}$ successful navigation trials, the average navigation steps is defined as 
\begin{align}
\bar{M}_{\rm s}= \frac{1}{N_{\rm s}}\sum_{m=1}^{N_{\rm s}}M_{\rm s}(m),
\end{align}
where $M_{\rm s}(m)$ is the number of navigation steps at the $m$th successful navigation trial.

\subsubsection{Safety Parameter ($S$)} For a successful navigation of $M_{\rm s}$ steps at locations ${\bf r}_1, {\bf r}_2, \ldots, {\bf r}_{M_{\rm s}}$ with a total number of $N$ unique detected obstacles during navigation, the safety parameter is the average minimum distance that the agent maintained from the detected obstacles during the navigation and is defined as  
\begin{equation}
    S \triangleq \frac{\sum_{m=1}^{N}\rho_{{\rm min}}(\textbf{r}^{m}_{\rm o})}{N},
    \label{eq:Average_Distance_Obstacles}
\end{equation}
where $\rho_{{\rm min}}(\textbf{r}^{m}_{\rm o}) \triangleq \min \big{\{} \lVert\textbf{r}_{1}-\textbf{r}^{m}_{\rm o}\rVert_{2}, 
\lVert\textbf{r}_{2}-\textbf{r}^{m}_{\rm o}\rVert_{2},
\ldots, \\ \lVert\textbf{r}_{M_{\rm s}}-\textbf{r}^{m}_{\rm o}\rVert_{2}\big{\}}$ with $\textbf{r}^{m}_{\rm o}$ is the location of the $m$th obstacle detected throughout the navigation. 

\subsubsection{Algorithm Complexity ($T_{\rm a}$)} The computational complexity of the algorithms is evaluated in terms of average run time of the algorithm for $N_{\rm m}$ Monte Carlo trials, and it is denoted by $T_{\rm a}$. 


\subsection{Simulation Results}
Figure \ref{fig:navigation_results} shows an example of a successful navigation of the agent employing the proposed ${\text{CR-BAPF}}^*$ algorithm. An unsuccessful navigation where the agent gets stuck in a local minima is also shown in Figure  \ref{fig:navigation_results}. As expected, by increasing the density of the obstacles in the navigation region, the chance of successful navigation decreases. 

\begin{figure}
    \centering
    \includegraphics[width=0.5\textwidth]{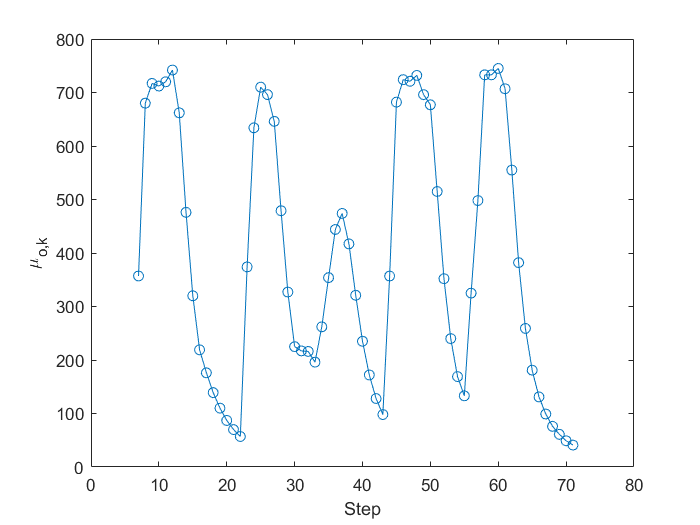}
    \caption{Optimized $\mu_{{\rm o},k}$ for the selected bacteria point versus navigation step for the proposed A-BAPF algorithm for a single successful run.}
    \label{fig:Optimized_mu_o}
\end{figure}

Figure \ref{fig:Optimized_mu_o} illustrates the optimized value of $\mu_{{\rm o},k} \in [1:1000]$ for the selected bacteria point versus navigation step for the proposed A-BAPF algorithm in a single successful run. As seen, there are sharp changes in the curve because the environment the agent is moving in is a cluttered environment and thus the number of detected obstacles, the distance from the detected obstacles, and the distance from the target significantly changes during the navigation. Our simulation result show that the A-BAPF algorithm can achieve a navigation success rate of more than $95\%$.

\begin{figure}
    \centering
    \includegraphics[width=0.5\textwidth]{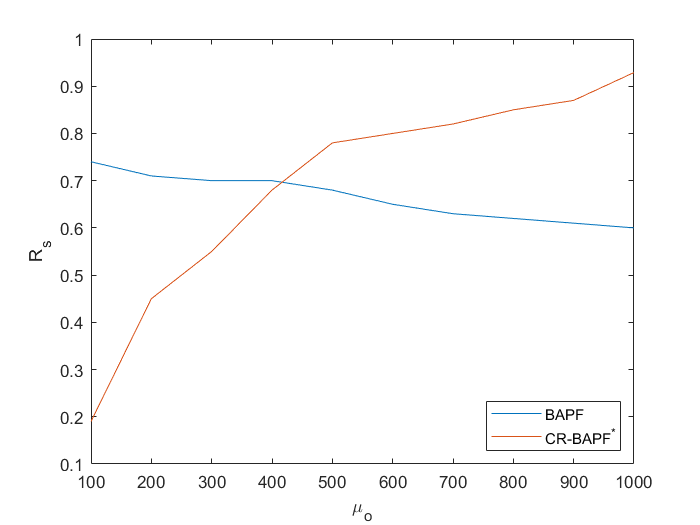}
    \caption{Rate of success $R_{\rm s}$ versus $\mu_{\rm o}$ for the proposed CR-BAPF\textsuperscript{*} and the BAPF algorithms.
    }
    \label{fig:Navigation Success Rates vs Fixed mu_o Values for IBAPF}
\end{figure}

Figure \ref{fig:Navigation Success Rates vs Fixed mu_o Values for IBAPF} compares the rate of success $R_{\rm s}$ versus $\mu_{\rm o}$ for the proposed ${{\text{CR-BAPF}}^*}$ and the BAPF algorithm. As seen, the ${{\text{CR-BAPF}}^*}$ algorithm outperforms the BAPF algorithm for $\mu_{\rm o}\in [416.67, 1000]$. The reason is that as $\mu_{\rm o}$ increases, the repulsive potential from the obstacles decreases. In the BAPF algorithm, the repulsive potential depends  only on the distance between the obstacle and the agent. When $\mu_{\rm o}$ increases, the risk of the agent getting close to an obstacle and collision increase. However, in the ${{\text{CR-BAPF}}^*}$ algorithm there are defined perimeters around each obstacle which control the repulsive potential value and prevent the agent from getting dangerously close to an obstacle. 
On the other hand, for $\mu_{\rm o}\in [1, 416.67)$, the repulsive potential is higher for the ${{\text{CR-BAPF}}^*}$. This results in a  higher success rate for the BAPF algorithm  because it becomes more likely for the ${{\text{CR-BAPF}}^*}$ algorithm to reach a local minima due to the high regional repulsive potential around each obstacle.


The comparison between the CAPF, the BAPF, and the proposed CR-BAPF and the CR-BAPF\textsuperscript{*} algorithms for different performance metrics including $R_{\rm s}$, $\bar{M}_{\rm s}$, $S$, and $T_{\rm a}$ is shown in 
Table \ref{table:comparison} for different obstacle densities. As seen, the proposed CR-BAPF\textsuperscript{*} algorithm outperforms the other algorithms in terms of success rate $R_{\rm s}$. The reason is that the proposed 
branching repulsive potential function and implementation of the random walk reduces the local minima trap and the chance of collision. Moreover, it can be seen that the success rate decreases as the number of obstacles in the environment increases which is a straightforward result. In terms of average navigation steps $\bar{M}_{\rm s}$, the number of steps generally increases for all methods when the number of obstacles in the environment increases. All the four algorithms perform well when it comes to the safety criteria $S$ because they offer $S>2$ m overall.
The CR-BAPF\textsuperscript{*} algorithm has a longer execution time compared to the others due to the random walk implementation which imposes extra steps to get out of a local minima but it yields a much higher success rate.



\begin{table}[!t] 
\renewcommand{\arraystretch}{1.4}
\caption{\bf Performance comparison between the four APF methods for different obstacle densities}
\centering
\begin{subtable}{\linewidth}\centering
\caption{$N \in {\mathcal{U}}{[20,45]}$}
\label{table:comparison_a}
\begin{tabular}{ |p{0.2\linewidth}||p{0.14\linewidth}|p{0.14\linewidth}|p{0.14\linewidth}|p{0.13\linewidth}| }
 \hline
 Algorithm & $R_{\rm s}$ & $\bar{M}_{\rm s}$ & $S$ (m)  & $T_{\rm a}$ (ms) \\ 
 \hline\hline
 CAPF \cite{Mohanta2016} & 0.333 & 91.26 & 2.71 & 26.80 \\ 
 \hline
 BAPF \cite{Ahmad2020} & 0.739 & 68.25 & 2.37 & 26.30 \\
 \hline
 CR-BAPF & 0.770 & 68.39 & 2.38 & 27.05\\
 \hline
 CR-BAPF\textsuperscript{*} & \textbf{0.935} & 70.47 & 2.35 & 31.63 \\ 
 \hline
\end{tabular}
\vspace{5pt}
\end{subtable}

\begin{subtable}{\linewidth}\centering
\caption{$N \in {\mathcal{U}}{[45,70]}$}
\label{table:comparison_b}
\begin{tabular}{ |p{0.2\linewidth}||p{0.14\linewidth}|p{0.14\linewidth}|p{0.14\linewidth}|p{0.13\linewidth}| }
 \hline
 Algorithm & $R_{\rm s}$ & $\bar{M}_{\rm s}$ & $S$ (m)  & $T_{\rm a}$ (ms) \\ 
 \hline\hline
 CAPF \cite{Mohanta2016} & 0.170 & 108.49 & 3.22 & 52.72 \\ 
 \hline
 BAPF \cite{Ahmad2020} & 0.552 & 77.82 & 2.40 & 66.87 \\
 \hline
 CR-BAPF & 0.490 & 69.24 & 2.42 & 44.18\\
 \hline
 CR-BAPF\textsuperscript{*} & \textbf{0.873} & 76.54 & 2.38 & 65.95 \\ 
 \hline
\end{tabular}
\vspace{5pt}
\end{subtable}

\begin{subtable}{\linewidth}\centering
\caption{$N \in {\mathcal{U}}{[70,95]}$}
\label{table:comparison_c}
\begin{tabular}{ |p{0.2\linewidth}||p{0.14\linewidth}|p{0.14\linewidth}|p{0.14\linewidth}|p{0.13\linewidth}| }
 \hline
 Algorithm & $R_{\rm s}$ & $\bar{M}_{\rm s}$ & $S$ (m)  & $T_{\rm a}$ (ms) \\ 
 \hline\hline
 CAPF \cite{Mohanta2016} & 0.157 & 111.93 & 3.53 & 71.64 \\ 
 \hline
 BAPF \cite{Ahmad2020} & 0.407 & 70.62 & 2.43 & 59.84 \\
 \hline
 CR-BAPF & 0.270 & 70.22 & 2.52 & 60.43\\
 \hline
 CR-BAPF\textsuperscript{*} & \textbf{0.812} & 83.36 & 2.44 & 108.65 \\ 
 \hline
\end{tabular}
\end{subtable}
\label{table:comparison}
\end{table}

\section{Summary}
We proposed the CR-BAPF algorithm to solve the local minima trap problem in the classical APF algorithms. The CR-BAPF takes the advantage of a branching repulsive potential function, which resulted in a navigation success rate of more than $75\%$ in lightly cluttered environments. This value increased to more than $90\%$ when the CR-BAPF algorithm was combined with the random walk technique (CR-BAPF\textsuperscript{*}). 
Our simulation results showed that 
the BAPF, the CR-BAPF, and the CR-BAPF\textsuperscript{*} algorithms all take less navigation steps to converge compared to the CAPF algorithm. Furthermore, the simulation results showed that all four algorithms maintain a good safety distance from the obstacles and that the corresponding execution times of the algorithms are acceptable. With the implementation of the random walk, the CR-BAPF\textsuperscript{*} outperforms the other algorithms at the expense of a higher execution time. In addition to the CR-BAPF algorithm, we also proposed the A-BAPF algorithm that is an adaptive version of the BAPF algorithm and achieves the upper bound performance of the BAPF. In the future, we aim to implement these algorithms in field agents to obtain experimental validation of the proposed solutions.

\section{Acknowledgements}
Mosab Diab is partially funded by the TU Delft \text{\textbar} Global Initiative. Mostafa Mohammadkarimi and R.T.Rajan are partially funded by the European Leadership Joint Undertaking (ECSEL JU), under grant agreement No. 876019, the ADACORSA (Airborne Data Collection on Resilient System Architectures) project. The code developed is provided in a repository at \url{https://github.com/M-DIAB/APF}.



\bibliographystyle{IEEEtran}
\bibliography{main.bib}


\thebiography
\begin{biographywithpic}
{Mosab Diab}{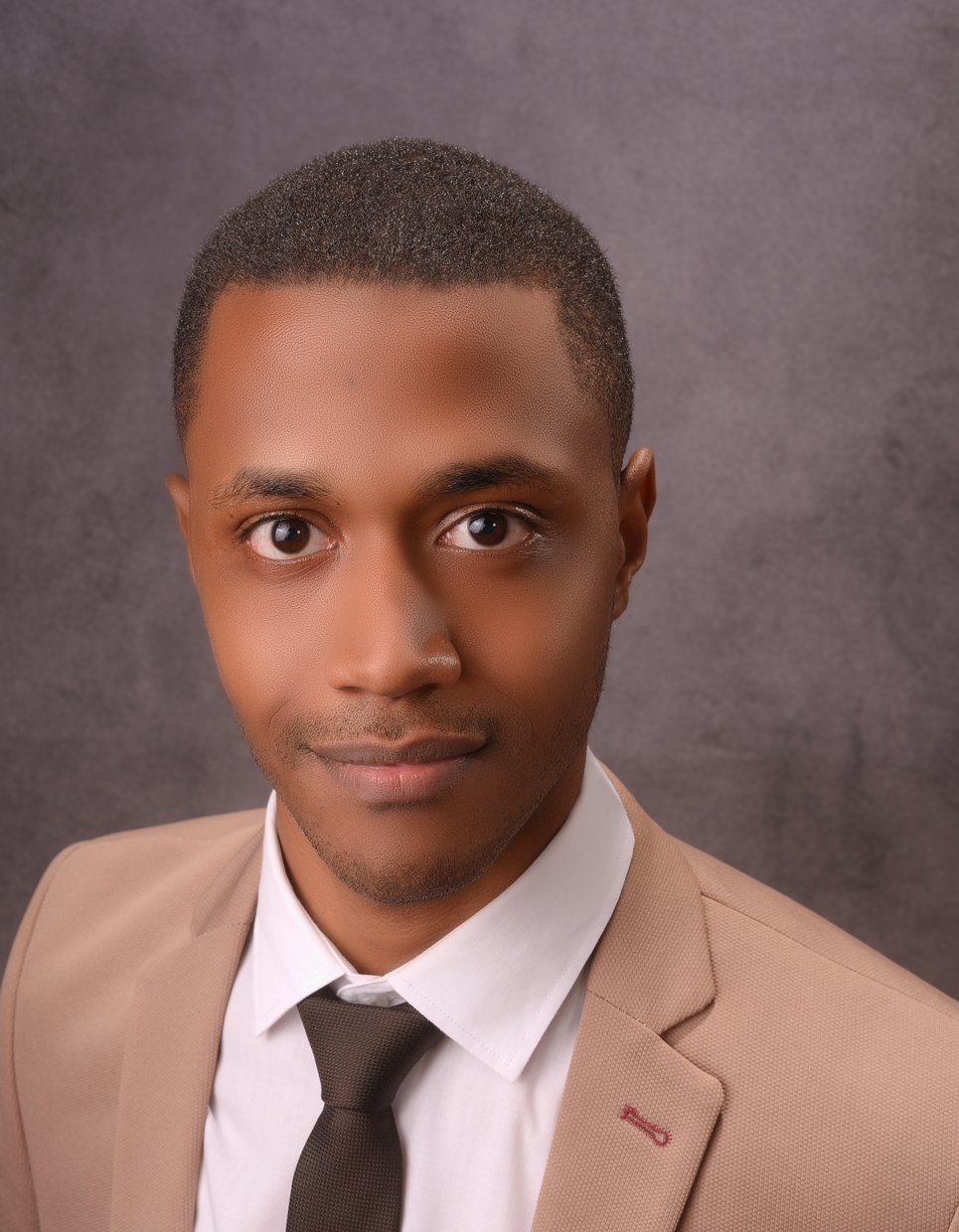}
received his B.Sc. degree from the University of Khartoum in Sudan in 2018 and his M.Sc. degree from the Delft University of Technology (TUD) in the Netherlands in 2022. Developed an interest for control systems and robotics early on, which culminated in him conducting research on myoelectric prosthetic hands for his bachelor's graduation project at the Center for Neuroscience and Biomedical Engineering at the University of Electro-Communications in Tokyo, Japan. For his master's thesis, he worked on path planning algorithms for autonomous agents in unknown cluttered environments.
\end{biographywithpic} 


\begin{biographywithpic}
{Mostafa Mohammadkarimi}{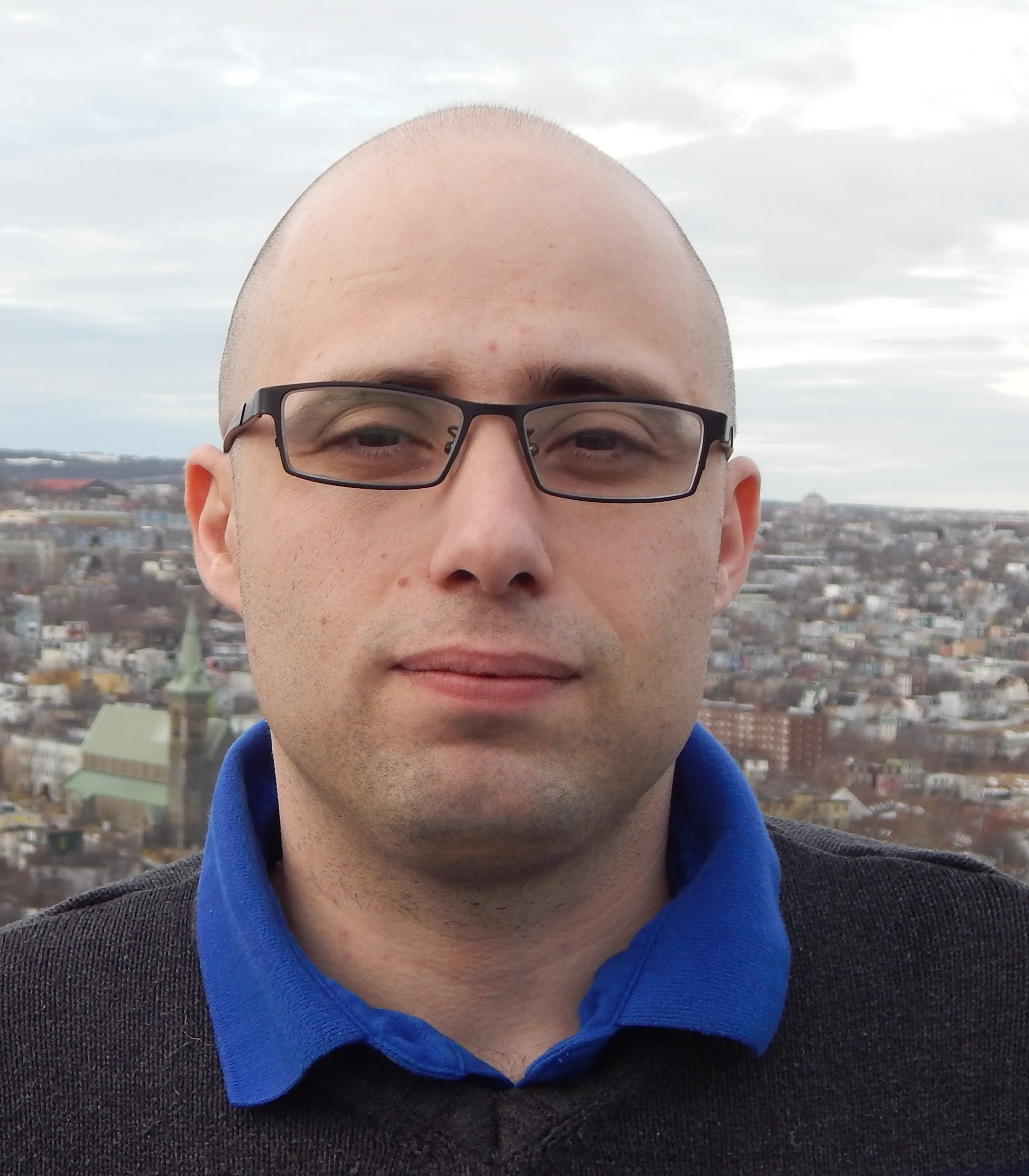}
received the M.Sc. degree in electrical engineering from K. N. Toosi University of Technology, Tehran, Iran, in 2011, and the Ph.D. degree in computer engineering from Memorial University, Canada, in 2017. From March 2016 to April 2017, he was a visiting Ph.D. student at Massachusetts Institute of Technology (MIT). From February 2018 to June 2021, he was a Post-Doctoral Fellow with the University of Alberta, Edmonton, AB, Canada, The University of British Columbia, Vancouver, BC, Canada,  Friedrich-Alexander-Universität Erlangen-Nürnberg, Erlangen, Germany. Since September 2021, he has been with the Signal Processing Systems Group, Department of Microelectronics, Delft University of Technology (TUD), Delft, The Netherlands. His research interests include wireless communications, statistical signal processing information theory, and localization $\&$ navigation. 

\end{biographywithpic}

\begin{biographywithpic}
{Raj Thilak Rajan}{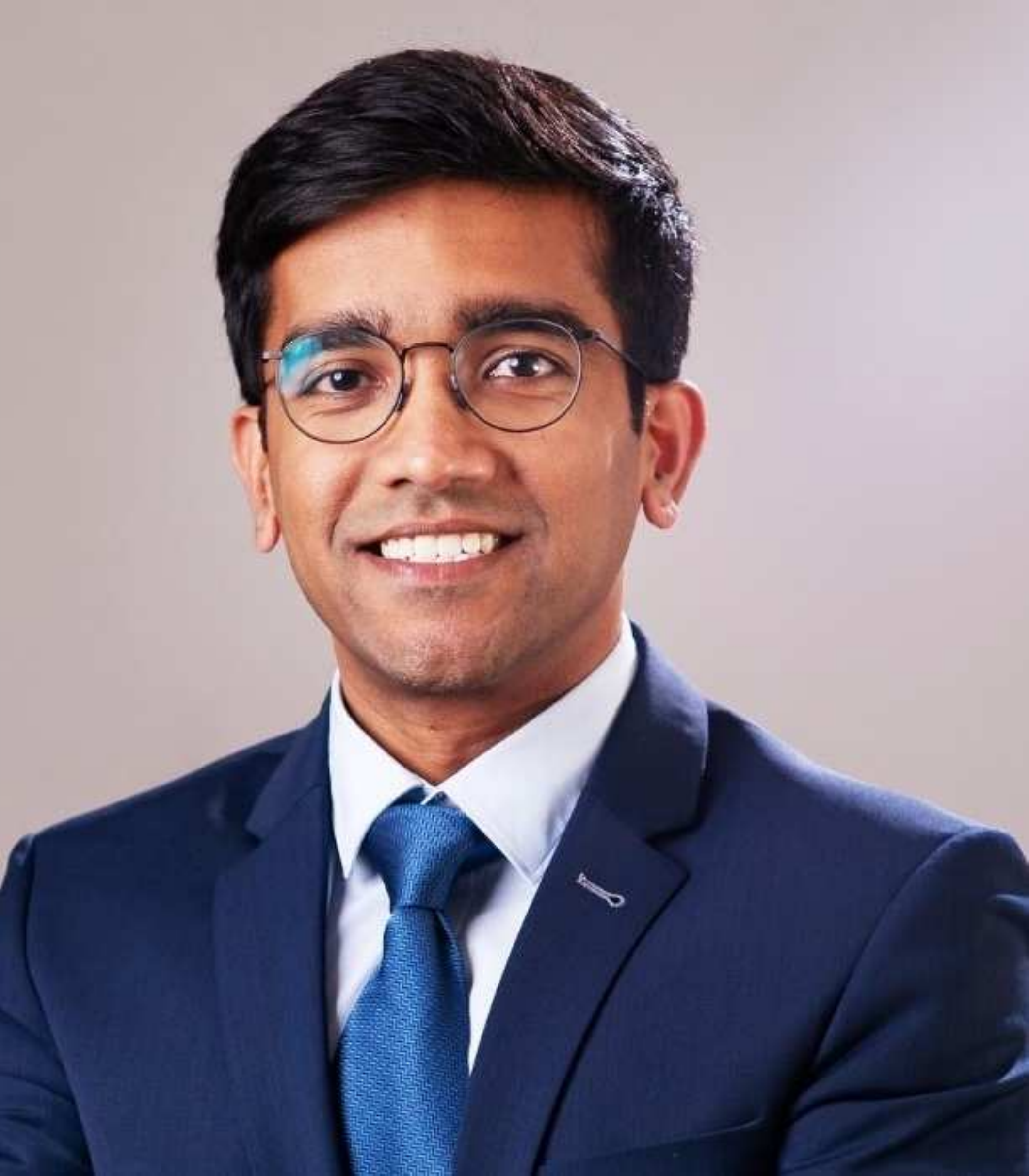} (S'11, M'17, SM'22) is an Assistant Professor with the Signal Processing Systems (SPS) group at the Faculty of Electrical Engineering, Mathematics and Computer Science (EEMCS) in the Delft University of Technology (TUD), and the Co-director of the Delft Sensor AI Lab. He is an elected member of the IAF SCAN (Space communications and Navigation) committee, the IEEE ASI (Autonomous Systems Initiative), and is the Associate Editor of the IEEE OJSP (Open Journal on Signal Processing).  His research interests lie in statistical machine learning and distributed optimisation, with applications to autonomous systems.
\end{biographywithpic}

\end{document}